# Statistical Analysis of Spatial and Temporal Variability of Maximum Precipitation Events on the Rio Grande do Sul

**Cleber Souza Corrêa\***

Instituto de Aeronáutica e Espaço - São José dos Campos/SP – Brazil

**Abstract**: A statistical analysis of precipitation at Rio Grande do Sul State was presented in this article. The aim of this work was to identify spatial and temporal patterns of maximum precipitation, which was achieved by fitting a theoretical variogram in maximum annual rainfalls and its times of occurrence. In the literature, it was found that this pattern occurs according to phenomena typical from middle latitude, such as low and high level jet, and interactions between them. Some years ago, the relationship between maximum annual rainfalls and synoptic predominant configurations was found. Therefore, this work sought on understanding the climatic characteristics that are important in Aerospace and Aeronautics, as extreme weather can cause numerous consequences in these activities. The use and validation of this proposed method would make possible its application in other regions of interest in the Brazilian Aerospace. Understanding these climatological features of the atmospheric circulation dynamics, and analyzing maximum annual rainfall would allow a more efficient and appropriate climate trend forecast and its application in aerospace activities.

## INTRODUCTION

In Rio Grande do Sul State, the formation of mesoscale convective systems (MCS) may be associated with different spatial and temporal scales, both in the planetary boundary layer (PBL) and mesoscale, with a wide range of mechanisms responsible for its formation.

Helfand and Schubert (1995) argue that low-level jets (LLJ) are the key processes in the transportation balance of moisture entering the mainland United States overnight.

Higgins et al. (1997) featured a daily cycle of precipitation using hourly observations from 1963 to 1993, revealing a well-defined maximum night on the Great Plains, in spring and in summer.

During the summer, there was 25% more precipitation during the night than during the day. The impact of LLJ on the balance sheet total precipitation shows an increase of more than 45% in average at night. In summer, this increase is considerably enhanced by the transport of water vapor from the Gulf of Mexico to the central part of the United

States of America (Whiteman et al., 1997). The examination of radiosonde data in cases with and without jet has a different specific humidity average of about 8.27 g/kg. The transport performed of the LLJ overnight associated with magnitudes of the specific humidity of about 10 g/kg can generate a very intense flux and severe precipitation. Chen et al. (1991) studied cyclones in the lee of the mountains (shortwave) in East Asia, in the Eastern Tibetan Plateau, especially in the summer rainy season over Southeast China and Taiwan area, from May to June.

This period is characterized by severe rainfall and is accompanied by the development and intensification of LLJ Northward in the lower troposphere. Chen and Yu (1988), analyzing data of precipitation during the rainy season on China and Japan, found 35 cases of daily rainfall around 100 mm per day. These severe rainfalls are closely associated with LLJ.

The authors found a likelihood of 84% with LLJ, with intensities of the order of 12.5 ms$^{-1}$, which occurred at the level of 700 hPa and 12 hours in advance of the heavy rainfall events.

When LLJ were present on the island of Taiwan, a probability estimate had a likelihood of 91% between precipitation events and the LLJ.

Lackmann (2002) discusses the contribution that the adiabatic potential vorticity produced in the maximum intensity of LLJ. This potential link between the bands of front precipitation and adiabatic redistribution of potential vorticity in the horizontal structure of the winds that accompany the LLJ, acting directly on the intensity of these. Maddox and Doswell III (1992) showed that the PBL may produce heat by convection and by condensation, and heat advection in the lower troposphere.

This process is associated with the circulations of the type Streams/LLJ, which perform that transport of sensible and latent heat in the lower atmosphere, dominating the differential vorticity advection in the middle troposphere, forcing upward vertically, resulting in the organization of convective events and generating MCS.

Uccellini and Johnson (1979) show that the high level jet (HLJ) and LLJ can often be coupled by the adjustments of the vertical mass, which occurs with the spread of HLJ.

This transport in the lower troposphere helps creating a favorable environment for the development of severe thunderstorms, especially when the interaction of the jets occurs through the intersection of the axes perpendicular occurring within the region's output of HLJ.

In South America, the work of Abdoulaev *et al.* (1996) showed that the MCS may be responsible for a significant portion of precipitation that occurs in Southern Brazil, occurring in at least 13 events per year, with intense rainfall and may be associated with LLJ intense.

Abdoulaev *et al.* (2001), studying nonlinear systems with mesoscale severe convection, observed the occurrence of a high rainfall probability exceeding 15 mm/h at about 6 o'clock in the morning, with a seasonal variability between the beginning and end of the convective cells. In summer, on average, the precipitations have a cycle of 6 to 30 hours and most MCS end life between 18 and 20 hours. This led Abdoulaev *et al.* (2001) to conclude that the cycle of night and early morning storms are amplified and modulated by the convergence of water vapor, which is mainly characterized by the LLJ.

In Rio Grande do Sul, Corrêa (2005) observed in radiosonde data held in Porto Alegre and Uruguaiana the dynamic structure of Streams/LLJ in the transport process at low levels.

Flows in the vertical wind profile can be defined as streams that have varying speeds of the order of 5 ms$^{-1}$ to LLJ ones (about 10 ms$^{-1}$), and they do not also present the vertical structure of the jet. The flows are significant because they may be associated with the convergence of mass and energy even with lower magnitude speeds than a LLJ acting on synoptic scales typical of mesoalpha ($\alpha$) and mesobeta ($\beta$), corresponding to spatial scales between 100 and 10 km.

Flows and LLJ in the vertical wind profile in low levels in the troposphere have different spatial and temporal scales, featuring a wealth of combinations of structures in PBL and the lower atmosphere, connecting with meso to continental scale. This dynamic structure at low atmosphere levels results in a layered structure, in which there is both its influence at an intermediate level (850 hPa) as the lowest level of around 950 hPa for convective storms over Rio Grande do Sul. An important effect of LLJ is the daily cycle of moisture as water vapor, and the convergence of transport within the PBL implies a kind of moisture storage over large areas and basins. These can generate meteorological phenomena, such as restricted visibility by haze and fog and cloud layers and type as low stratus and stratucumulus. In cases where there is a positive balance, this moisture storage becomes a source of water vapor for future convection occurrence. Therefore, this structure daily transport plays an extremely important role in water balance of a watershed in middle latitudes.

There are few scientific studies linking the LLJ events with high rainfall, and the majority of studies on this topic are case ones with reanalysis data of global models, which according to Berbery and Collini (2000) limits the exploitation of results due to low resolution of the data.

De Barros and Oyama (2010) analyzed the weather systems associated with the occurrence of precipitation in the Alcântara launch base between 2005 and 2006. It aimed at characterizing the systems associated with precipitation events in the Alcântara Launch Center (CLA) between 2005 and 2006, using total hours of precipitation, NCEP reanalysis data of NCEP/NCAR, brightness temperature of the satellite GOES-12, and long-wave radiation. To this end, we defined criteria to identify the weather systems of large, meso and local scale precipitation associated with the CLA, the results were that 60% of the precipitation processes were stratiformed derived and 40% convective.

This paper attempts to test the methodology of Geostatistics, calculating the semi-variogram of maximum rainfall, and using it in future researches to the region of Alcântara, in Maranhão. The use of data observed in this study, by calculating the semi-variogram of maximum precipitation, is an attempt to identify spatial and temporal patterns in precipitation maximum.

## DATA AND METHODOLOGY

Flows of water vapor are in the vertical profile of wind and LLJ direction and they have variable frequency, are associated with weather systems, and may be responsible for generating part of the convection and, hence, rainfall. These precipitations vary in time and space, as well as their maximum.

To study the relationship between the LLJ and the maximum precipitation, precipitation data at 52 rainfall stations of the Agência Nacional de Águas (ANA) in the state of Rio Grande do Sul were analyzed. We attempted to use only observation stations with complete series.

It was calculated, using the statistical software GENSTAT© maximum rainfall in one, two, three, four and five days for each station, comprising the years from 1971 to 2000. Also, the day when the event of maximum precipitation happened was determined, followed by a procedure similar to that used for the events for one, two, three, four and five days, for each station.

This statistical analysis used a model implemented in GENSTAT© that has the method of the semi-variogram, allowing a quantitative representation within the range of a localized phenomenon (Huijbregts, 1975), which can be expressed as Eq. 1:

$$Z(x) = \mu + \varepsilon(x), \tag{1}$$

where,

$Z(x)$ is the value of the random variable,

$\mu$ is the mean of the variable $Z(x)$,

and the term $\varepsilon(x)$ is a self-correlated random function with zero mean and variance defined by Eq. 2

$$\text{var}[\varepsilon(x) - \varepsilon(x+h)] = E[\{\varepsilon(x) - \varepsilon(x+h)\}^2], \tag{2}$$

and $h$ is the distance between the point $x$ and $x + h$. It is assumed that, on average, $Z$ is constant in space such as Eq. 3:

$$E[Z(x) - Z(x+h)] = 0. \tag{3}$$

The degree of dependence between points $x$ and $x + h$ is represented by the variogram $2\gamma(h)$, which is defined as the mathematical expectation of the square of the difference between the values given by Eq. 4:

$$\text{var}[Z(x) - Z(x+h)] = E[\{z(x) - Z(x+h)\}]^2 = 2\gamma(h) \tag{4}$$

The variance depends not only on the distance $h$ and the position $x$. The amount is $\gamma$ semi-variance, which in turn is a function of $m$ that is used to estimate the variogram. It is assumed the hypothesis "Matheron" and also that the process is second order stationary, in which the average random process is generally constant, that is, the expected value $E[Z(x)] = \mu$. It is also assumed that the variance is constant ($\text{var}[Z(x)] = \sigma 2$), the covariance exists and it is only dependent on the difference between $Z(x)$ and $Z(x + h)$, represented by Eq. 5 and 6:

$$C(h) = E[\{Z(x) - \mu\} \{Z(x+h) - \mu\}], \tag{5}$$

and

$$C(0) = \text{var}[Z(x)] = E[\{Z(x) - \mu\}^2] \tag{6}$$

The covariance is correlated by the semi-variogram by Eq. 7:

$$\gamma(h) = C(0) - C(h). \tag{7}$$

To calculate the variogram, the "FVARIOGRAM" function is used, which creates a semi-variogram experimental set of variable values $Z(x)$ distributed into one or two dimensions, using the formula defined by Eq. 8:

$$\hat{\gamma} = \frac{1}{2m(h)} \sum_{i=1}^{m(h)} \{Z(x_i) - Z(x_i+h)\}^2 \tag{8}$$

where:

$Z(xi)$ and $Z(xi + h)$ are the values of the position $xi + h$. The term $m(h)$ is the number of pairs for comparison, which contributes to the estimate. As the data show scattering irregular spatial precipitation, $h$ is discretized in such a way that its value is constant, the variation in the distance increase and the same change of direction are considered to be isotropic. The "FVARIOGRAM" function calculates the experimental semi-variogram. With the estimated semi-variogram already done, we have tried to fit a mathematical model for the semi-variogram obtained.

In this analysis, we used a semi-linear variogram, by using the "MVARIOGRAM" in GENSTAT©. Table 1 shows an example of the estimate of the calculation of the semi-variogram in a period of a year, based on 50 observation posts and a maximum spacing of 57m, and a length of 0.25 was considered when calculating the semi-variogram, homogeneity in all directions. Where $N$ is the number of observations, $h$ is the length and $V$ is the estimated variance of the semi-variogram (Table 1).

Table 1. Estimations of computing the semi-variogram.

| N | h | V |
|---|---|---|
| 10 | 0.165 | 6750.6000 |
| 27 | 0.396 | 6715.9259 |
| 54 | 0.617 | 6687.6944 |
| 56 | 0.895 | 9888.2232 |
| 60 | 1.135 | 7201.1333 |
| 66 | 1.363 | 7861.8106 |
| 69 | 1.623 | 8264.5000 |
| 64 | 1.882 | 8418.9297 |
| 62 | 2.121 | 7393.1452 |
| 53 | 2.381 | 7048.8396 |
| 61 | 2.624 | 9089.8525 |
| 78 | 2.869 | 8451.1282 |
| 70 | 3.142 | 11893.2429 |
| 93 | 3.368 | 13544.8978 |
| 84 | 3.626 | 14686.9583 |
| 71 | 3.878 | 15791.6127 |
| 73 | 4.130 | 12657.9932 |
| 60 | 4.388 | 15485.4917 |
| 48 | 4.629 | 15836.8021 |
| 25 | 4.857 | 17130.5800 |

After this calculation is performed, an adjustment of a linear estimate of the semi-variogram, obtaining an estimate of the percentage of variance explained by the model, can be seen in the linear fit of the model in Fig. 1.

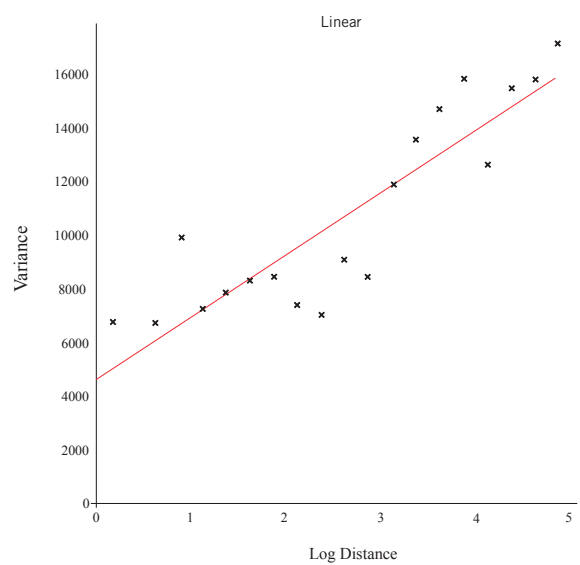

Figure 1. Adjustment of the linear model to the semi-variogram.

This analysis uses a statistical hypothesis that performs the calculation of the ratio of the variance, showing how the model explains data variance. Thus, a variance of about 100% would be interpreted as resulting from a strong linear relationship, which can be interpreted as absence of spatial or temporal consistency of precipitation.

Moreover, close to zero (0%) would be typical of a low-linear relationship, which could be interpreted as the absence of spatially and temporal uniform precipitation. As the theoretical model has a linear adjustment, it is observed in the calculation of linear regression model that the best adjustment explains most of the variance of the spatial or temporal maximum precipitations.

The method was used to perform a linear adjustment of the semi-variogram. In this type of analysis, more complex nonlinear models as model "boundedlinear" and Gaussian could be used. However, it was used the linear model, which although simple has good results. This analysis also tried to consider the spatial homogeneity of variance, however this could not be considered, which would bring the calculation of the semi-variogram with estimates of predominant directions. The annual results of the variance explained by the linear model for each year, between 1971 and 2000, are compared with the signal of El Niño/Southern Oscillation (ENSO) obtained on the website www.cpc.ncep.noaa.gov, with the aim of correlating changes in scale with the spatial and temporal trends of the maximum precipitation.

Figure 2 shows the spatial distribution of jobs and Table 2 shows the 52 rain gauge stations of the National Water Agency (ANA) on the state of Rio Grande do Sul.

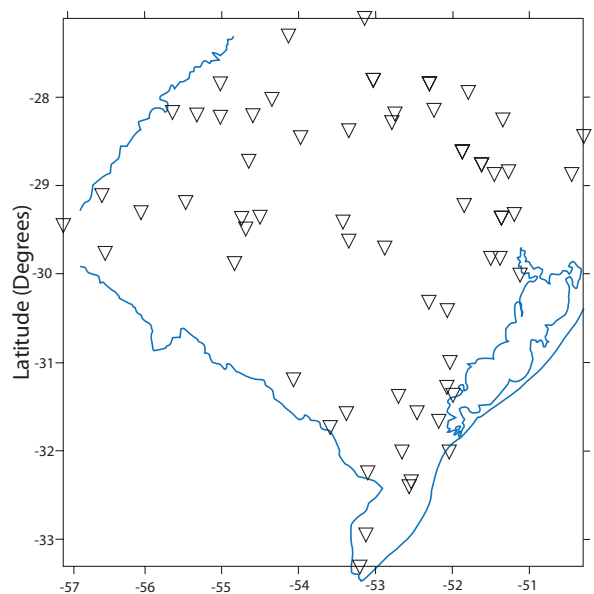

Figure 2. Spatial distribution map of rainfall and weather stations in Rio Grande do Sul.

Table 2. Rainfall stations of ANA in Rio Grande do Sul.

| Code | Locality | Latitude | Longitude | Network |
|------|----------|----------|-----------|---------|
| 2850009 | Passo Tainhas | -28.87 | -50.45 | ANA |
| 2851003 | Antonio Prado | -28.85 | -51.28 | ANA |
| 2851021 | Passo do Prata | -28.87 | -51.45 | ANA |
| 2851022 | Passo Migliavaca | -28.62 | -51.87 | ANA |
| 2851024 | Prata | -28.77 | -51.62 | ANA |
| 2951010 | Encantado | -29.23 | -51.85 | ANA |
| 2951022 | Nova Palmira | -29.33 | -51.19 | ANA |
| 2951024 | Porto Garibaldi | -29.82 | -51.38 | ANA |
| 2951027 | São Vendelino | -59.19 | -51.37 | ANA |
| 2951028 | Sapucaia do Sul | -29.82 | -51.50 | ANA |
| 3051005 | Guaiba Country Club | -30.11 | -51.65 | ANA |
| 3052011 | Quiteria | -30.42 | -52.07 | ANA |
| 3151003 | São Lourenço do Sul | -31.37 | -51.99 | ANA |
| 3152002 | Bouqueirão | -31.28 | -52.08 | ANA |
| 3152011 | Passo do Mendonça | -31.00 | -52.05 | ANA |
| 2853014 | Santa Clara do Ingai | -28.73 | -53.19 | ANA |
| 2952003 | Botucarai | -29.72 | -52.89 | ANA |
| 2953008 | Dona Francisca | -29.63 | -53.35 | ANA |
| 3152003 | Canguçu | -31.39 | -52.70 | ANA |
| 3152008 | Granja São Pedro | -31.67 | -52.18 | ANA |
| 3152016 | Ponte Cordeiro de Farias | -31.57 | -52.46 | ANA |
| 3153007 | Pedras Altas | -31.74 | -53.59 | ANA |
| 3153008 | Pinheiro Machado | -31.58 | -53.38 | ANA |
| 3252005 | Granja Coronel Pedro Osório | -32.01 | -52.65 | ANA |
| 3252006 | Granja Cerrito | -32.35 | -52.54 | ANA |
| 3252008 | Granja Santa Maria | -32.40 | -52.56 | ANA |
| 3253001 | Arroio Grande | -32.24 | -53.09 | ANA |
| 3253003 | Granja Osório | -32.95 | -53.12 | ANA |
| 3253004 | Herval | -32.03 | -53.40 | ANA |
| 2751006 | Paim Filho | -27.70 | -51.77 | ANA |
| 2751007 | Sananduva | -27.95 | -51.81 | ANA |
| 2752006 | Erebango | -27.85 | -52.30 | ANA |
| 2753004 | Linha Cescon | -27.81 | -53.03 | ANA |
| 2754001 | Alto Uruguai | -27.30 | -54.13 | ANA |
| 2755001 | Porto Lucena | -27.85 | -55.02 | ANA |
| 2852006 | Carazinho | -28.29 | -52.79 | ANA |
| 2852007 | Colonia Xadrez | -28.19 | -52.75 | ANA |
| 2854003 | Giruá | -28.03 | -54.34 | ANA |
| 2853003 | Conceição | -28.46 | -53.97 | ANA |
| 2854005 | Passo Major Zeferino | -28.73 | -54.65 | ANA |
| 2854006 | Passo Viola | -28.21 | -54.60 | ANA |
| 2855001 | Garruchos | -28.18 | -55.64 | ANA |
| 2855002 | Passo do Sarmento | -28.21 | -55.32 | ANA |
| 2855005 | Fazenda S. Cecília de Butui | -29.00 | -55.68 | ANA |
| 2956005 | Itaqui | -29.12 | -56.56 | ANA |
| 2954001 | Cacequi | -29.88 | -54.82 | ANA |
| 2954004 | Ernesto Alves | -29.37 | -54.73 | ANA |
| 2954005 | Furnas do Segredo | -29.36 | -54.50 | ANA |
| 2954007 | Jaguari | -29.49 | -54.69 | ANA |
| 2955002 | Cachoeira Santa Cecília | -29.20 | -55.47 | ANA |
| 2956006 | Passo Mariano Pinto | -29.31 | -56.05 | ANA |
| 2956007 | Plano Alto | -29.77 | -56.52 | ANA |

## RESULTS

Table 3 shows the results of the linear fit of the variogram models, calculated by presenting their respective percentages of variance explained by the linear model. The results show the existence of specific spatial and temporal coherence of the maximum rainfall in some years in the period between 1971 and 2000 on the state of Rio Grande do Sul.

Note that in this period there were four years with significant coherence, spatial and temporal, 1974, 1978, 1985 and 1997. These years had an explained variance with more meaningful values for intensities and scales of two, three, and four days. Such characteristics show weather systems, which cycles have a tendency of being timescale of the order of one week.

During the formation of a weather system with these features, in the timescale of a week, there is a combination of different types of upright flow or LLJ, which shows maxima and minima patterns of behavior. They, when associated with the values of a certain magnitude of LLJ and maximum precipitation, are connected to a particular group of weather systems (atmospheric waves, and locking systems MCS), which are part of a cascade structure comprising the variable climate in this respect, when this occurs, these structures can generate possible in-homogeneities, both spatial and temporal precipitations.

Table 3. Percentage of variance explained by the adjustment of the linear model between 1971 to 2000 about Rio Grande do Sul State.

| Year | Number of Observations | Percentage (%) | | | | | | | | | |
|---|---|---|---|---|---|---|---|---|---|---|---|
| | | MaxI_01 | MaxI_02 | MaxI_03 | MaxI_04 | MaxI_05 | DiaM_01 | DiaM_02 | DiaM_03 | DiaM_04 | DiaM_05 |
| 1971 | 50 | 0 | 18 | 33 | 30 | 24 | 75 | 71 | 70 | 59 | 81 |
| 1972 | 52 | 38 | 59 | 57 | 57 | 63 | 0 | 0 | 0 | 16 | 0 |
| 1973 | 50 | 42 | 53 | 47 | 57 | 48 | 32 | 36 | 32 | 25 | 0 |
| 1974 | 51 | 55 | 43 | 47 | 46 | 45 | 50 | 72 | 56 | 52 | 34 |
| 1975 | 50 | 35 | 39 | 39 | 40 | 43 | 49 | 71 | 26 | 33 | 35 |
| 1976 | 51 | 15 | 20 | 20 | 32 | 32 | 7 | 0 | 3 | 33 | 0 |
| 1977 | 49 | 16 | 0 | 19 | 0 | 14 | 11 | 20 | 18 | 0 | 0 |
| 1978 | 52 | 42 | 65 | 74 | 79 | 76 | 0 | 18 | 70 | 78 | 67 |
| 1979 | 50 | 78 | 84 | 80 | 87 | 87 | 15 | 0 | 13 | 33 | 7 |
| 1980 | 52 | 0 | 4 | 0 | 0 | 0 | 0 | 0 | 2 | 0 | 0 |
| 1981 | 51 | 0 | 0 | 14 | 25 | 56 | 0 | 15 | 0 | 0 | 0 |
| 1982 | 49 | 50 | 63 | 67 | 63 | 66 | 0 | 26 | 37 | 30 | 20 |
| 1983 | 50 | 74 | 53 | 8 | 0 | 5 | 0 | 0 | 38 | 21 | 43 |
| 1984 | 49 | 18 | 8 | 0 | 0 | 0 | 0 | 0 | 2 | 23 | 15 |
| 1985 | 49 | 57 | 75 | 62 | 69 | 45 | 18 | 42 | 45 | 79 | 50 |
| 1986 | 48 | 17 | 49 | 41 | 43 | 58 | 0 | 0 | 18 | 7 | 0 |
| 1987 | 49 | 67 | 83 | 79 | 80 | 79 | 33 | 0 | 0 | 3 | 0 |
| 1988 | 49 | 37 | 41 | 46 | 42 | 50 | 3 | 0 | 4 | 28 | 29 |
| 1989 | 52 | 27 | 0 | 1 | 22 | 48 | 4 | 0 | 4 | 0 | 0 |
| 1990 | 46 | 0 | 0 | 0 | 0 | 0 | 15 | 0 | 0 | 12 | 10 |
| 1991 | 42 | 0 | 0 | 13 | 26 | 33 | 33 | 32 | 19 | 5 | 9 |
| 1992 | 47 | 0 | 0 | 0 | 0 | 0 | 0 | 0 | 10 | 0 | 1 |
| 1993 | 45 | 2 | 0 | 0 | 7 | 16 | 0 | 0 | 0 | 0 | 0 |
| 1994 | 46 | 9 | 0 | 14 | 0 | 10 | 0 | 0 | 1 | 0 | 9 |
| 1995 | 44 | 27 | 46 | 66 | 62 | 50 | 3 | 20 | 10 | 0 | 0 |
| 1996 | 42 | 5 | 9 | 44 | 33 | 31 | 0 | 8 | 7 | 0 | 0 |
| 1997 | 32 | 36 | 25 | 68 | 68 | 61 | 18 | 15 | 52 | 28 | 46 |
| 1998 | 47 | 0 | 0 | 2 | 0 | 0 | 0 | 12 | 64 | 56 | 88 |
| 1999 | 45 | 35 | 53 | 15 | 57 | 46 | 11 | 12 | 35 | 17 | 0 |
| 2000 | 41 | 0 | 68 | 71 | 56 | 55 | 0 | 0 | 0 | 0 | 12 |
| Average | 48 | 26 | 32 | 34 | 36 | 38 | 13 | 16 | 21 | 21 | 19 |
| SD | 4 | 24 | 29 | 28 | 28 | 26 | 19 | 22 | 23 | 24 | 26 |

According to Table 3 for 1972, 1973, 1979, 1982, 1983, 1986, 1987, 1995, 1999 and 2000, there is spatial coherence in the maximum rainfall, but temporal coherence was not observed.

Table 4 shows that El Niño had two predominant situations, the predominance of spatial coherence years (five years) and years without spatial and temporal coherences (five years). In La Niña years, there were three situations, namely: three years with temporal coherence, three years without spatial coherence, and temporal and spatial coherence in four years.

The years 1974, 1978, 1985 and 1997 showed spatial and temporal coherence and they are also from El Niño, La Niña, and ENSO unsigned.

The spatial coherence years were divided into years with El Niño and La Niña, but the El Niño years have been characterized by two predominant situations, with a spatial coherence and the other without spatial and temporal coherences. The weather systems of these periods had a dynamic, in which there was the predominance of certain specific meteorological scales generating spatial coherence.

In El Niño years, the HLJ on South America exhibit greater intensity and persistence than in other phases of ENSO, and, consequently, a greater intensity of LLJ in these periods. The passage of frontal systems, such as short-waves, mesoscale convective and other weather systems with organized convective structures tend to be characterized with maximum spatial coherence, but these weather systems were not the same as the standard for consistency over time, which can be seen in El Niño years in Tables 3 and 4, Nicolini and Saulo (2000), Liebmann *et al*. (2004), and Saulo *et al*. (2007).

The physical processes that are involved in the generation of precipitations as in MCS have different spatial dimensions, and their formation is associated with different meteorological mechanisms, instabilities forming isolating or forming groups, as in cells *Cumulunimbus* (Cb) (a few miles) to clusters Cb in a mesoscale convective complex (MCC), with tens to hundreds of kilometers. Therefore, the maximum rainfall in space may vary from a few to hundreds of kilometers.

This is an important feature of the precipitation, since the generation mechanisms of MCS not always repeat in time,

have different sizes and life cycles. The existence of climate variability in the state of Rio Grande do Sul implies that rainfall varies in the four seasons, both in intensity and duration and in area of coverage. Rao and Hada (1990) demonstrated the existence of significant correlation between precipitation in the spring in Rio Grande do Sul and the ENSO signal during the same season or earlier, with the possibility of predicting seasonal rainfall.

Table 4. Comparison of the percentage of variance explained by the adjustment of the linear model with the ENSO signal, between 1971 to 2000.

| Year | El Niño | La Niña | No Sign of ENSO |
|------|---------|---------|-----------------|
| 1971 |         | 2       |                 |
| 1972 | 3       |         |                 |
| 1973 |         | 3       |                 |
| 1974 |         | 4       |                 |
| 1975 |         | 2       |                 |
| 1976 |         |         | 1               |
| 1977 | 1       |         |                 |
| 1978 |         |         | 4               |
| 1979 |         |         | 3               |
| 1980 |         |         | 1               |
| 1981 |         |         | 1               |
| 1982 | 3       |         |                 |
| 1983 | 3       |         |                 |
| 1984 |         | 1       |                 |
| 1985 |         | 4       |                 |
| 1986 | 3       |         |                 |
| 1987 | 3       |         |                 |
| 1988 |         | 1       |                 |
| 1989 |         | 1       |                 |
| 1990 |         |         | 1               |
| 1991 | 1       |         |                 |
| 1992 | 1       |         |                 |
| 1993 | 1       |         |                 |
| 1994 | 1       |         |                 |
| 1995 |         | 3       |                 |
| 1996 |         | 1       |                 |
| 1997 | 4       |         |                 |
| 1998 |         | 2       |                 |
| 1999 |         | 3       |                 |
| 2000 |         | 3       |                 |

Data from ENSO. Source: www.cpc.ncep.noaa.gov.
1: no consistent spatial / temporal; 2: with temporal coherence;
3: with spatial coherence; 4: with consistent spatial/temporal.

The maximum precipitation in 1971, 1975, and 1998 showed no spatial coherence, but showed consistency over time. This fact is interesting because it was La Niña years (1971 and 1998 – and 1975 moderate to strong), the climatic situation characterizes periods of drought, causing a tendency to decrease frequency and intensity of weather phenomena with different scales on Rio Grande do Sul.

Another important point illustrated in Table 3 shows the situation in which the lack of spatial and temporal coherence, namely, 1976, 1977, 1980, 1981, 1984, 1988, 1989, 1990, 1991, 1992, 1993, 1994, 1996, is distributed in all stages of ENOS signal. These periods have not showed characteristics of spatial and temporal coherences, which can result in characterizing the physical process in the generation of maximum precipitation of years. It shows the existence of a wide variety of weather and switching scales of different types of weather structures in space and time.

The different types of weather systems had relations in space, and time variations can occur only at mesoscale, or act in concert with other scales (meso to meso, meso and macro, macro to macro scale), resulting in changes in the characteristics of weather systems, and may intensify them or diminish them in intensity.

The combinations of these structures show a trend of stability and uniformity in growth, development, and persistence of weather systems, hence, an increase of spatial coherence, but these same relationships may or may not have temporal coherence. An example of this is when areas are formed by divergent instabilities associated with HLJ, they may have irregularities on a spatial region of the order of mesoscale.

It is defined MaxI_01, MaxI_02, MaxI_03, MaxI_04, and MaxI_05, with a maximum of one day of rain and the maximum that occurred in two, three, four, and five days. The DiaM_01, DiaM_02, DiaM_03, DiaM_04 DiaM_05 are defined as the days of the year in which these events maximum precipitation occurred, thus examined the temporal variability.

When this divergence associated with HLJ is intensified and deepened, generating a wave structure in the atmosphere (tropical cyclones) with a higher spatial and temporal scales, the level of continental scale leads to a dynamic structure and a possibly more homogeneous field of precipitation. In El Niño years, the HLJ are more intense and persistent and they are associated with LLJ configurations that have a tendency to spatial homogeneity.

The physical mechanisms of weather systems are strongly baroclinic (turbulent), however, through their relationships between the different meteorological scales involved, their physical properties in space and time change, both intensify and weaken these systems. There is, within this dynamic system, a range of interaction between different synoptic structures, which is very important in the maintenance and generation of baroclinic and convective processes. This range is characterized by flows that occur in the PBL, which is the maximum transport in the vertical wind profile of a structure with LLJ. The role of LLJ within this turbulent structure is to be a link between the different spatial scales of meso/meso and meso/continental scales.

In this situation, in each spatial scale, maximum rainfalls are associated. Therefore, in heavy rainfall periods, they are associated with large-scale meteorological vertical development in the atmosphere, whether as locks, associated with the configuration of HLJ, whether as vortices cold at altitude in the atmosphere and the years of ENSO (El Niño) with strong LLJ. This structure of continental scale and meso, with strong LLJ, can generate maximum precipitation with regular distribution in space and time, the size of the spatial and temporal scale involved, as the year of 1997.

In other situations, the LLJ may be weaker and predominant in the order and mesoscale, which can generate and maintain structures and baroclinic instabilities that can cause irregular or regular precipitations in space and time. In this context, within this dynamic structure described, the flow or LLJ always occurs, independent of whether there is spatial homogeneity or temporal variability of rainfall.

Therefore, the same weather scale may act differently in the generation of maximum rainfall, featuring a very complex structure that comprises a set of dynamic weather systems, and which, together, form the field of precipitation and is influenced by local factors, with regional and large scales.

## CONCLUSIONS

In this study, it was observed that in La Niña years, there is no predominant feature that characterizes an ENSO signal defined spatial or temporal coherence, characterizing the absence of trend for the maximum annual rainfall in those years. However, during the El Niño, there was well-characterized evidence for the spatial coherence tendency. Therefore, the use of Geostatistical analysis calculating the semi-variogram and fitting a linear model, provided information of the spatial and temporal coherences of rainfall on the state of Rio Grande do Sul. These results could obtain characteristics of predominance of the maximum annual rainfall and their application to forecast climate trends. Consequently, it enabled its use in planning activities in aviation airports in Southern Brazil, and the estimation of the most likely year for the Aerospace activities.